# Voter Verification of BMD Ballots Is a Two-Part Question: Can They? Mostly, They Can. Do They? Mostly, They Don't

Philip Kortum, Michael D. Byrne, and Julie Whitmore

Rice University, Houston, Texas

**ABSTRACT**

The question of whether or not voters actually verify ballots produced by ballot marking devices (BMDs) is presently the subject of some controversy. Recent studies (e.g., Bernhard, et al. 2020) suggest the verification rate is low. What is not clear from previous research is whether this is more a result of voters being unable to do so accurately or whether this is because voters simply choose not to attempt verification in the first place. In order to understand this problem, we conducted an experiment in which 108 participants participated in a mock election where the BMD displayed the voters' true choices, but then changed a subset of those choices on the printed ballot. The design of the printed ballot, the length of the ballot, the number of changes that were made to the ballot, the location of those changes, and the instructions provided to the voters were manipulated as part of the experiment. Results indicated that of those voters who chose to examine the printed ballot, 76% detected anomalies, indicating that voters *can* reliably detect errors on their ballot if they will simply review it. This suggests that administrative remedies, rather than attempts to alter fundamental human perceptual capabilities, could be employed to encourage voters to check their ballots, which could prove as an effective countermeasure.

After the passage of the Help America Vote Act (U.S. Congress, 2002), voting officials quickly moved to utilize this federal money to purchase new voting systems that were purported to make it easier to vote while maintaining or enhancing voting security. These replacement systems were typically electronic voting systems, and while many of the systems did indeed make it easier to vote, particularly for those voters who needed accessibility accommodations (American Federation for the Blind 2002; Cross et al, 2009; National Federation of the Blind, 2020; Piner & Byrne, 2011), the promise of higher security was largely lost with these new systems. Indeed, there are numerous examples of electronic voting systems that were so far below the generally accepted standards of electronic security as to endanger the integrity of votes that were cast on them (Balzarotti et al., 2010; Bishop, 2007; Epstein, 2015; Feldman, Halderman, & Felten, 2007; Proebstel et al., 2007). As these machines have begun to show their age, municipalities are now moving to implement voting methods that have far better security characteristics than the first generation of electronic voting devices.

Although end-to-end cryptographic computer voting systems are capable of ensuring security and auditability in a voting system (Benaloh et al, 2012; Chaum et al, 2008; Popoveniuc & Vora, 2008), there is growing consensus that some form of paper trail is needed to provide the highest levels of security in auditability of the voting process (Norden & McCadney, 2019; Berger et al, 2018). Indeed, 37 of the 50 states in the US currently require some form of paper record of the vote (NCSL, 2019). This paper record can take three general forms: 1) a paper ballot that can be hand marked by the voter, who then deposits it into a either a secure ballot box for later counting, (either by hand or by an electronic counting machine), or a scanner for immediate counting 2) an electronic ballot marking device (BMD), which uses a computer interface to gather voter selections and then print those selections onto a paper ballot. The paper ballot is the official voting record, and the ballot marking device does not retain any of the voting information. Counting the ballots takes place either by hand or using an electronic counting machine, and 3) A hybrid of these two systems, where an electronic voting machine is used to mark a ballot which is stored and counted electronically, but the machine creates a paper ballot as well. The paper ballot serves as a backup to the electronic record, and allows for auditing the electronic record. These systems are commonly known as VVPAT voting systems (voter verifiable paper audit trails).

Because of the cost effectiveness and relative ease of implementing a paper-only voting system, many municipalities who had used HAVA monies to purchase their voting machines have begun to move back to paper ballots as these systems have aged out (Hautala, 2018; NSCL, 2019). Since 2006, the number of precincts voting with paper has increased approximately 30% (Pew Research, 2016) and the trend seems to be continuing. On the surface, switching to paper seems to be an effective and efficient method for conducting elections, given its lengthy history and the ease of understanding its implementation. However, there are some issues that make the use of paper ballots alone difficult. Notably, in large voting districts where there are hundreds of precincts, printing and managing this multitude of ballots can be difficult. Further, the use of paper-only ballots creates significant, if not insurmountable, difficulties for voters with disabilities.

BMD and VVPAT devices are seen as a reasonable way around the limitations of paper-only elections. The multitude of precinct ballots—including multiple languages, if necessary—are stored electronically and generated on the spot for a voter from any precinct (an especially important characteristic in cities where vote centers are being utilized). Such systems can straightforwardly support those voters with physical or visual disabilities, utilizing any of the accommodative technologies that are widely available in the marketplace. From a security standpoint, a voter verifiable paper trail is generated so that voting officials have a physical record of the vote (whether it serves as the primary record or as a backup), which can be used to conduct post-election audits.

However, it has been argued that ballot marking devices do not have any significant security advantage over voting systems where a computer is used alone to record the ballot choices made by the voter (Appel, DeMillo, & Stark, 2020). The claim is made that malicious software on the ballot marking device could display the voter's correct selections on the computer review screen, but then substitute its own choices on the printed ballot. The voter, believing that the display on the BMD represents their selections, may not carefully check the paper record, and so fail to capture the act of cheating by the BMD. Several studies (Acemyan & Kortum, 2013; Campbell & Byrne, 2009; Everett, 2007) have demonstrated that significant portions of a review screen can indeed be modified from the voters intentions, and that voters often do not notice these deviations from their intentions.

The concern about the security risk of compromised BMDs tainting an election has resulted in some security experts advocating for the banning of BMD devices altogether, and relying on hand marked paper ballots exclusively. For example, a lawsuit in Georgia has been filed to prohibit the state government from implementing new BMD systems, rather than paper-only systems in the upcoming election (Fowler, 2019). In addition, some voting security experts have expressed serious concerns about California's certification of new VSAP (Voting Solutions for All People) ballot marking device in Los Angeles County (Bajak, 2020).

The evidence showing that voters are not reliable detectors of altered races on a review screen or a printed ballot appears strong. Previous research (Acemyan & Kortum, 2013; Bernhard et al, 2020; Campbell & Byrne, 2009; Everett, 2007) shows a relatively low rate of detection of malicious changes, and this has led many in the voting community to suggest that people are simply not good at this kind of checking task, and therefore cannot be counted on to be a vital component in the security of voting systems.

Of note, however, is that the data reported from these studies has typically examined the percentage of voters—across all of the voters who participated in the study—who detect a change on a ballot or review screen that has been maliciously altered. Thinking about the data in this way obscures the true performance of those voters who actually *did* check their ballot for malicious changes. Voters who did not even look at their ballot are counted as having detected no changes, even though they made no attempt to do so.

A different, more appropriate. way of examining these data would be to look at the detection rate among those voters who *actually review* their ballot. This changes several things about the approach. First, differences in outcomes also have different policy implications. If the problem is that voters who examine the ballot cannot find discrepancies, then this implicates the difficulty of the task and perhaps the design of the ballot. Solutions here are likely to be more difficult, as performance on this task is less likely to be under volitional control by voters.

On the other hand, if voters are generally successful at detection when it is attempted, but the rate of attempt is low, then the problem might be addressable by more traditional policy interventions such as voter education campaigns and poll worker instructions. The problem would then switch from trying to get voters to be better checkers (which can be difficult due to cognitive and perceptual limitations of the human in this task) to simply getting more voters to actually review their ballot.

In this new framework of understanding the data, the solution is as much a policy problem as it is a human factors problem. Voting officials could focus on administrative solutions that could help encourage or require voters to take the time to review their ballot, while psychologists could focus on what kinds of interface elements might aid voters when they actually do check their ballot. Are there things that could be done that would help a voter in examining their printed ballot, once the decision has been made by that voter to check their ballot, so that it would be more probable that any malicious activity by the BMD would be caught? Wallach (2019) has demonstrated that not all voters have to catch a malicious BMD in action. A single catch by a single voter in a given precinct is enough to alert the authorities to isolate and forensically evaluate a machine or machines in that precinct. Therefore, even small improvements in the number of voters who detect malicious changes to their ballot could be significant in the improvement of BMD security.

What forms of improvements could be utilized in order to help improve the likelihood that voters will detect a malicious change to their ballot? First, creating printed ballots that have superior readability characteristics might be one way of doing this. Many of the BMD/VVPAT outputs are difficult to read because of the physical form that they take. Races are printed using suboptimal fonts, formatted in ways that make it difficult to easily read the ballot, and use paper that may be difficult to handle. Simple changes to text formatting, font selection, and paper choice might prove beneficial in helping voters in this respect.

Second, it could be that the length of the ballot has an impact on whether voters are able to accurately detect changes. Extremely long ballots may overwhelm the user and cause them to have reduced effectiveness in catching malicious changes to their ballot, while shorter ballots might result in better performance. If this were the case, then precincts could take steps to modify elections in ways that would support the best voter behavior in ballot checking, or by creating paper ballot trails that would engender some of those same characteristics (e.g., putting fewer races on each ballot page that a voter needs to check).

Of course, human factors specialists still have much to contribute to the critical first half of the problem as well, namely how do we get more voters to check their ballots in the first place? Is simply informing the users that they need to check their printed form of the ballots sufficient? Voters may not understand that validation of the printed record is important , or they may forget to do so in their rush to leave the polling station. Giving appropriate instructions on the importance of taking the time to examine a ballot at an appropriate juncture in the voting process might have substantial benefits in raising these detection rates, and the form, location, and timing of these admonitions would all be important to understand. Recent work by Bernhart et al. (2020) suggests that asking users to check their ballot may be an effective strategy for increasing the likelihood of catching a changed ballot.

The questions, then, are which of the two failure modes (failure to check, or failing when checking) dominate, and what factors contribute to success rates for both?

## METHOD

### Participants

A total of 108 people were recruited from the greater Houston area using advertisements and Craigslist. Eligible participants needed to be U.S. citizens, at least 18 years of age, speak English proficiently, have normal motor capability and normal or corrected-to-normal vision. Of the participants, 59 were female and 49 were male. Their ages ranged from 18 to 66 with a median age of 33 years. There was a diverse racial makeup, with a 52.3% identifying as African American, 23.9% as Caucasian, 13.8 % as Hispanic/Latino/Chicano/Mexican American, 4.6% as Asain American, 0.8% as American Indian, and 4.6% as other. Education levels were also representative, with 4.6% of participants indicating they had not graduated high school, 25.7% reporting a high school diploma or GED, 45% reporting some college or an associates degree, 15.6% reporting a bachelor's degree, and 9% having a graduate degree. All participants were compensated $20 for their participation.

### Design

The experiment was a 2 (design of printed ballot) x 2 (length of ballot) x 2 (number of changes) x 2 (location of changes) x 2 (instructions) between-subjects design.

*Design of printed ballot:* The interface of the electronic ballot was consistent throughout the experiment, but the printed paper ballot was a reproduction of traditional ES&S systems (Figure 1A) or the new VSAP system (Figure 1B). These two formats were chosen to determine if there were differences between a traditional design and one based on established user-centered design principles, respectively.

*Length of ballot:* Two ballot lengths were used in the study: a 40 contest ballot (long), and a 5 contest ballot (short).

*Number of changes:* Two extremes of vote flipping were examined: a single targeted flip or a large number of flips (40%).

*Location of changes:* The location of the changes made were in the beginning or the middle of the ballot. For one flip, the change was either the vote for President or a random contest in the middle. For 40% flips, the change was the vote for President and random contests in the beginning, or all random contests in the middle.

*Instructions:* Two instruction levels were examined: A maximum instruction (or "primed") condition, in which text reminders were printed at the beginning and end of the electronic ballot, as well as verbal reminders to the participants before they voted and after they printed the ballot (but before they submitted it to the ballot box). The verbal reminder informed participants that the printed ballot was the official ballot. There was also a no-instruction (or "no prime") condition, where no ballot checking instructions were included on the ballot, and no verbal reminders were given to the participants at any time during the experiment.

**Materials**

To simulate an electronic voting machine, an 18 inch touch screen tablet computer was used. The screen was mounted in a frame that provided the proper position and angle for voters. Qualtrics was used to build the electronic ballot and all interface designs for the ballot and electronic review screen resembled those of the VSAP system (Figure 2). Previous studies (Acemyan & Kortum, 2013; Campbell & Byrne, 2009; Everett, 2007) have used computer generated names on the ballots, but in order to more realistically represent an actual voting situation, universally recognizable (but non-political) people were used as candidates (with arbitrary party assignment). For example, names such as Mark Twain, Thomas Edison, and Bill Gates were used as candidates. Is this way, the names the participants were choosing would be as familiar to them as politicians running for office would be, which might aid in the detection of flipped votes, while still retaining some of the important beneficial characteristics of fake names identified by Quesenbery and Chisnell (2009).

A wizard-of-oz protocol was used for ballot printing. To the user, it appeared as though their ballot was being printed when they pressed the "print ballot" button. In reality, the experimenter would surreptitiously print a ballot of the correct form (design of printed ballot, length of ballot, number of flips, location of changes, and instruction mode) for the participant to examine and cast. This greatly simplified the running and coding of the experiment.

**Procedure**

Participants first completed an IRB-approved consent form. Next, they were given a list of candidates, and instructed to vote according to that slate. Then, they were randomly assigned a condition and (depending on the condition) given verbal instructions to check their printed ballot when they were done voting. Participants would then vote using the BMD. On the last screen they would be prompted to press the "print ballot" button and the experimenter would simultaneously print the compromised ballot. Depending on the condition, the experimenter would prompt the participant again to check their ballot before submitting it to the ballot box. The participants then cast their printed ballots in a well marked ballot box. After their ballot was submitted, the participants were given a survey that collected information about demographics, voting history, and usability.

Figure 1: A) ES&S-style printed ballot form. B) VSAP-style printed ballot form. Both ballots show the long ballot with checking instructions.

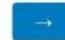

Figure 2: A representative vote selection screen on the ballot marking device.

## RESULTS

**Overview of Approach**

While the overall detection rate in this experiment was 17.6% (not particularly high), this number by itself is somewhat deceptive; the story is more complex than this. Prior views of anomaly detection, including our own, from the first studies by Everett (2007) to the most recent (Bernhard, et al. 2020), have taken the approach of looking directly at what variables influence overall detection rates, assuming all variables directly affect detection. However, we now believe this model of the process is oversimplified. In fact, in order for a voter to detect an anomaly on a printout, the voter has to do two things: (1) decide to examine the ballot, and (2) detect any changes during the examination. If the voter fails to examine the ballot, the voter will obviously fail at detection; these are thus distinct issues. The goal of this approach is to provide some separation between the question of "*do* voters detect anomalies?" and "*can* voters detect anomalies?" It matters little whether or not they are able to if they do not attempt to do so.

From a data analysis standpoint, this approach requires two separate analyses with somewhat different variables being used. Some factors, in principle, cannot affect voters' choice to examine the ballot, because they are only evident if the ballot is examined. The number and location of ballot changes fall into this category. On the other hand, urging voters to examine the ballot should only affect whether or not they do so. (It is possible that this could potentially affect accuracy when doing so, but this seems unlikely.) Other factors likely affect both the actual and perceived difficulty of the detection task, and thus may affect both voters' choice to examine and how well they do so. Ballot length and the design of the ballot are factors that might affect both.

While we do not have a perfect measure of whether or not our voters made a legitimate attempt to verify their ballot, we did measure whether or not a voter actually examined the printout, which is at least a necessary condition for verification. Thus, we have two analyses here, both done with logistic regression: one with examination as the dependent variable and instructions, ballot length, and ballot type as predictors; and one with detection as the dependent variable and ballot length, ballot type, change location, and change amount as predictors. The overall model is depicted in Figure 3. The sample size for the second analysis is obviously smaller (since not all voters examined the printout) and thus has lower statistical power.

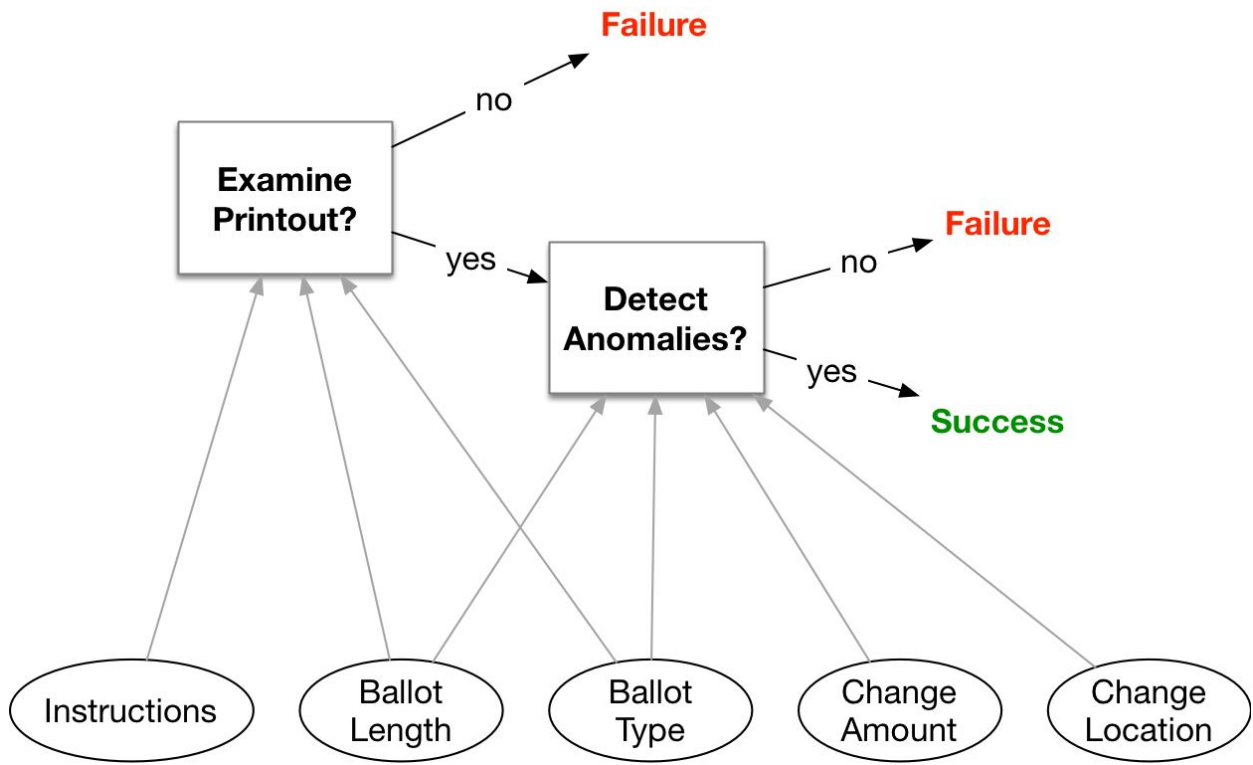

Figure 3. Model for anomaly detection. Voters who fail to examine the ballot obviously fail to detect anomalies. Voters who do examine the ballot may still fail. Different experimental factors may affect the choice to examine, detection performance, or both.

**Choice to Examine**

The largest cause of the overall low detection rate was that most of the voters in our experiment did not examine the ballot in the first place; only 25 of 108 voters (23.1%) did so. Two of the three manipulated variables significantly impacted the examination rate: instructions and ballot length. Descriptive statistics for instructions and ballot length appear in Figures 4 and 5 respectively. Instructions were a significant predictor (44% vs. 2%) for the logistic regression, $\chi^2 = 13.20$, $p < .001$ as was ballot length (33% vs. 14%), $\chi^2 = 5.9$, $p = .015$. The overall regression had a McFadden $r^2$ of 0.32 and produced a classification accuracy of 83.3%.

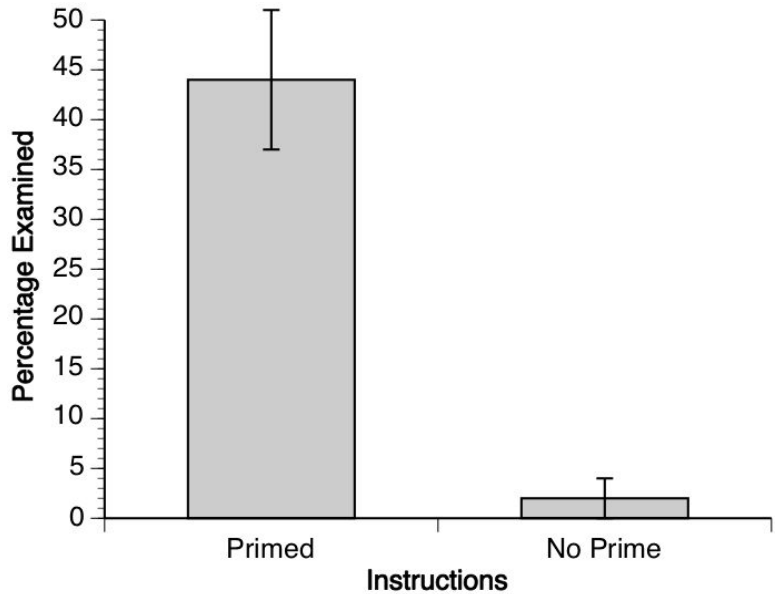

Figure 4. Percentage of voters who examined their ballot as a function of whether or not they were primed to do so by instructions and the poll worker. Error bars represent one standard error of the mean.

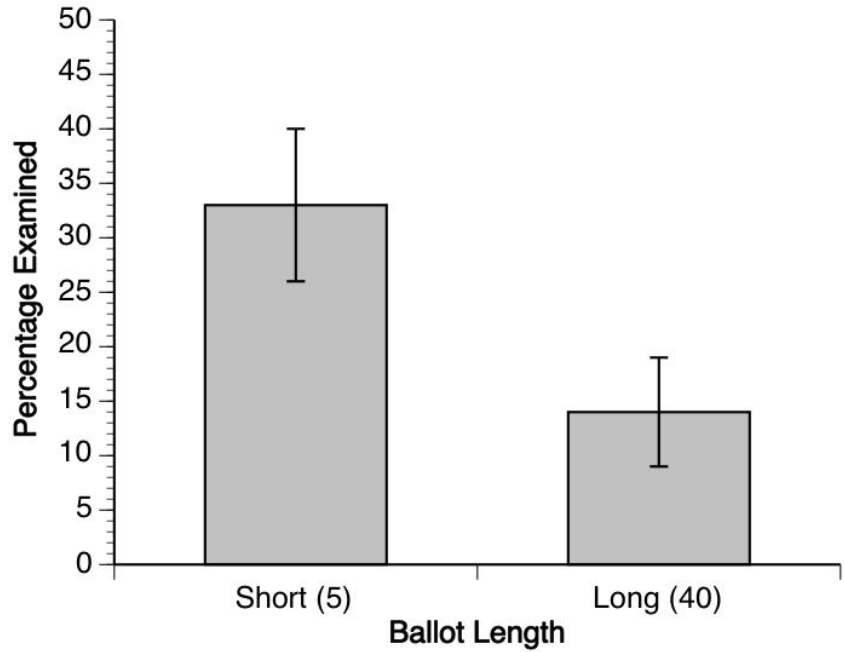

Figure 5. Percentage of voters who examined their ballot as a function of the length of the ballot. "Short" ballots had 5 contests, "long" ballots had 40 contests. Error bars represent one standard error of the mean.

There was no evidence that ballot type had any effect on the choice of whether or not to examine the ballot; both the VSAP-style printout and the ESS-style printout produced identical 23% examination rates.

**Detection Performance**

Unlike the decision to examine, actual detection performance was quite good. Of the 25 voters who actually examined the printout, 19 of them (76%) detected at least one anomaly. This suggests that many voters *can* find errors in a BMD printout; the problem is that not many in our experiment actually tried.

Descriptive results for ballot length, ballot type, change amount, and change location are presented in Figures 6 through 9. Unfortunately, none of the manipulated variables were statistically significant predictors (overall McFadden $r^2$ = 0.24 for the full regression; the best predictor was change amount, $\chi^2$ = 2.50, $p$ = .12). The problem is that statistical power with only 25 voters is limited, so despite large descriptive differences, results here are not conclusive.

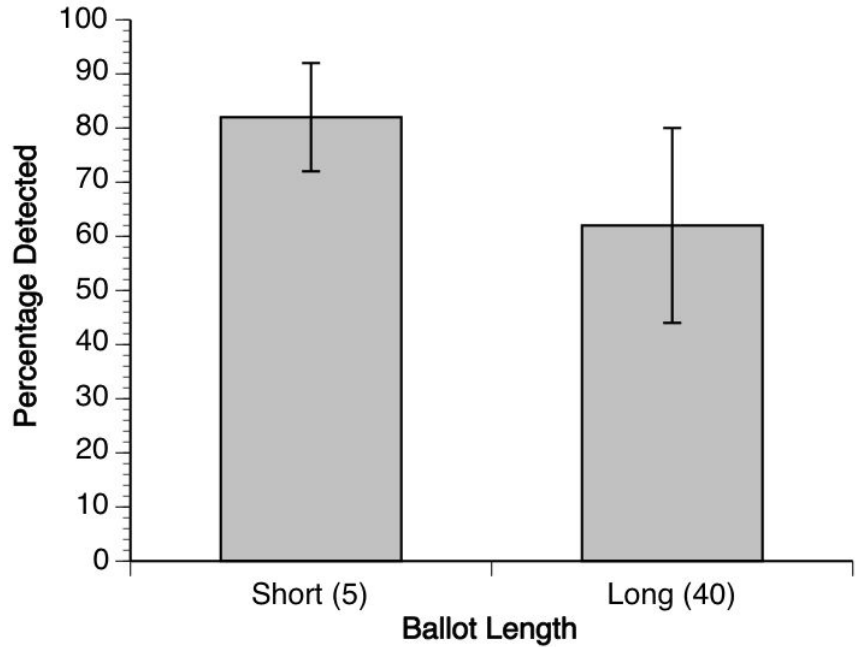

Figure 6. Percentage of voters who detected one or more anomalies as a function of the length of the ballot. "Short" ballots had 5 contests, "long" ballots had 40 contests. Error bars represent one standard error of the mean.

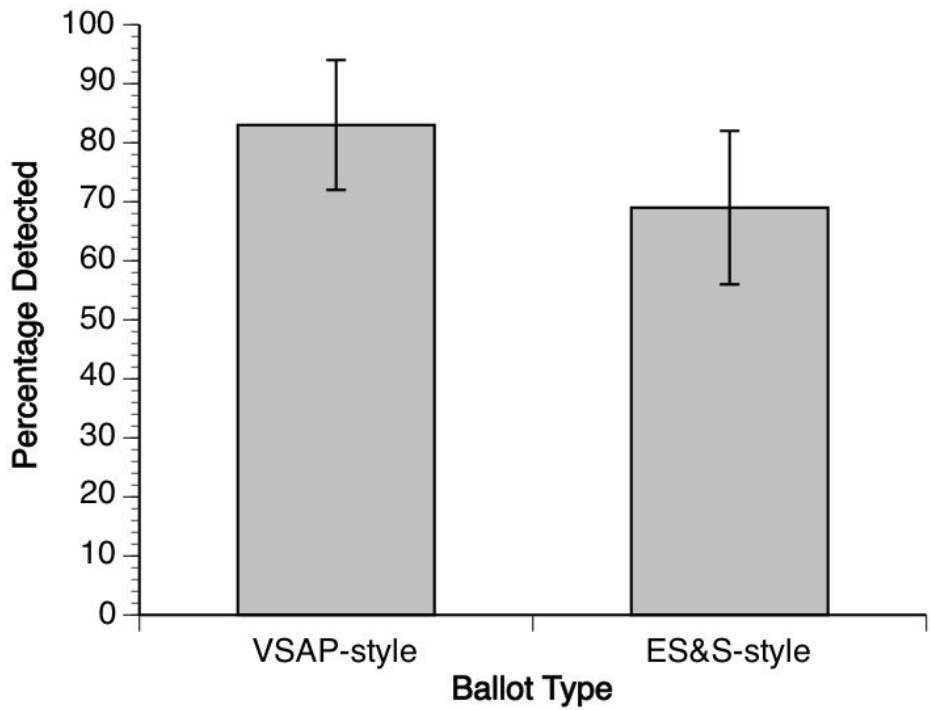

Figure 7. Percentage of voters who detected one or more anomalies as a function of which style of printout they received. Error bars represent one standard error of the mean.

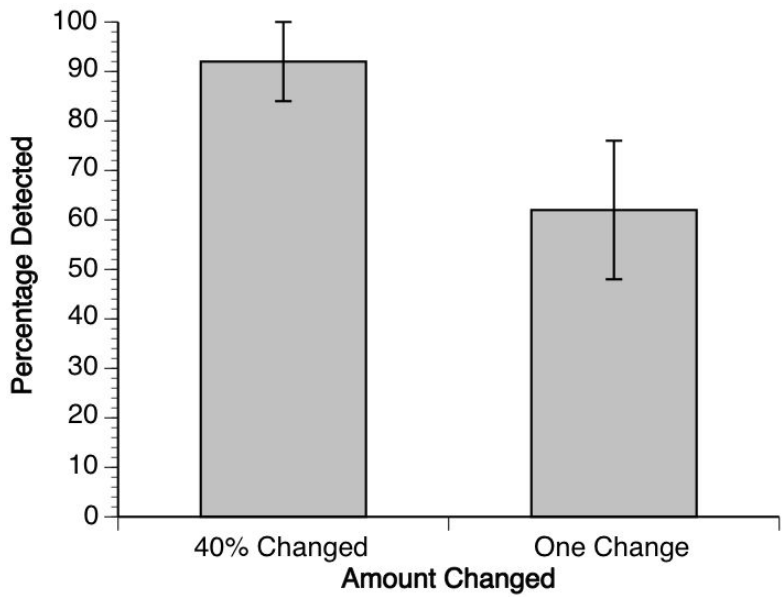

Figure 8. Percentage of voters who detected one or more anomalies as a function of how many votes were altered on the printout. Error bars represent one standard error of the mean.

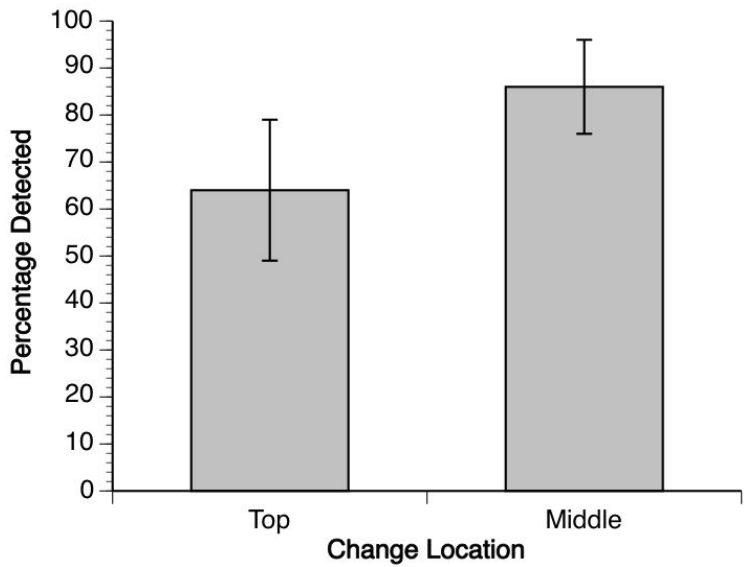

Figure 9. Percentage of voters who detected one or more anomalies as a function of where on the printout the anomalies occurred. Error bars represent one standard error of the mean.

Despite the differences not being statistically significant, it is worth noting that three of the four differences line up with the expected direction of difference. That is, more voters detected changes on short ballots, on the VSAP-style ballots, and when there were more changes to find.

Only the last result, that more voters detected anomalies when they were at the middle (vs. the top of the ballot), is surprising. More research along these lines—with a larger sample size—is warranted to determine whether or not these results are meaningful.

## DISCUSSION

Our new perspective on the detection problem, and the accompanying findings, have important implications for public policy and future research. First, the separation of choice to examine and detection performance, and the fact that the former was poor while the latter was good, suggests that this is likely a problem addressable by policy. It is not the case that people are intrinsically bad at this task and limitations in human performance have to be overcome. Instead, this appears to be primarily a problem of motivation and education. A substantial part of the risk can likely be mitigated through behavioral changes or through administrative changes to the process.

Our results are also likely underestimating true rates of examination. Studying human behavior in a voting context is difficult because in real elections, privacy is protected and we cannot, in principle, know a voter's actual intent. Because our experiment was not a real election, it is likely that participants were not as motivated to find and report ballot errors as they might be in an actual election where their vote was important, and could have an impact on the election outcome. For this reason, it is probable that the examination rates found in this study represent the lower bound, and that data from real elections would probably produce higher rates of examination, since voters would have significant incentive to ensure that their vote was cast correctly.

Note also that in this study, participants were given slates of candidates and instructed to vote that slate. They retained the slate throughout the entire process, meaning they could have used the slate in order to do a systematic check of their ballot. Since not all voters create lists and then vote using them, this experimental detail may have increased the number of detections that were made by voters who reviewed their ballots. That said, if additional research suggests that the use of a slate increases detection probability, as was suggested in studies by Bernhard et al (2020), then administrative and informational actions could be taken to encourage voters to bring and use sample ballots in the voting booth.

It is important to note that some of these issues are not necessarily unique to BMDs, and even cross into the realm of hand-marked paper ballots. Simply because a ballot is marked by hand does not imply that the ballot itself is prepared by hand. Ballots are generated using computers, and those computers are also subject to malicious and/or malfunctioning software. An attack on a computer connected to a ballot-on-demand printer may result in a system that produces erroneous ballots, perhaps missing races or with candidates' party affiliations switched. Guaranteeing election integrity requires officials or voters to notice such changes. Unlike BMD attacks, these are detectable after the fact, but if such a situation is detected later, such as in a post-election audit, what is the remediation? The best solution is to detect this prior to the voter completing a ballot, and this kind of detection presents a similar problem to detecting vote flips on a BMD.

Since our data demonstrate that people actually *can* detect changes in their ballots if they will simply take the time to check, then it seems that the next logical question is "What can be done to get people to take the time to examine their ballots in the first place?" There are a number of possibilities, all of which would require additional research in order to understand how efficacious they might be. Some of these possibilities are straightforward administrative actions that could be implemented with little cost and relatively minor changes to existing election protocols in the precinct.

For example, perhaps the solution is as simple as providing written instructions to the voters that they should check their ballot before they cast it, backed up by signage in the polling area, or admonitions on the ballot that it should be reviewed before being cast. Alternatively, or in concert with such interventions, poll workers might prove to be effective. Poll workers could stand by the ballot box and remind people to check their ballots before they cast them. The combination of social pressure and verbal instruction might prove to be all that is necessary to increase checking rates substantially.

Certainly, more unconventional ways of encouraging voters to check their ballots should also be considered as part of this extended research effort. For example, adding a location in the polling place that is between the ballot marking station and the ballot box, where voters would be "expected" to examine their ballot might be a solution that employs concepts from environmental psychology. Temporal strategies could also be employed, where voters would be given timeouts in order to check their ballot before it was cast.

Of course, if such efforts are successful in getting most or all of the voters to check their ballots, then we must also investigate how to effectively deal with people finding errors and making sure that those are viewed not as human mistakes but as warning signs of a potential malicious agent in a BMD.

It is still possible, however, that such efforts will not be successful and an insufficient proportion of voters will verify their ballots. Critics of BMDs have argued that it should not be the voters' job to verify that the voting equipment is working, and we have some sympathy with this position. Fortunately, the security situation here is different than with electronic machines that do not produce a paper record. Here, if a machine cheats or is defective, it immediately provides evidence that it is faulty, and thus can be caught. In addition to interventions designed to encourage voters to check their ballots, we recommend that election officials deploying BMDs perform live audits as outlined in Wallach (2019).

In this scenario, election officials would field teams whose sole purpose was to perform random audits of the BMDs to make sure that they were performing as expected. The strength of this approach is that any anomalies found can be directly attributed to a malfunctioning or malicious BMD, because the audit team would be following a script, and deviations from that script would necessarily indicate a problem. This would eliminate much of the uncertainty of voter-detected ballot changes, where suspected deviations would have to accumulate before the alarm could be reliably sounded.

Unfortunately there is a potential weakness in this strategy as well. If those who are trying to do harm to the election can determine when an auditor is using the BMD to perform an audit, then the malicious software can simply provide "correct" results in that particular instance. For example, if the audit team were to examine the machines right before the polls opened, all a malicious attacker would have to do would be to observe the system clock. If it noted that the voting was taking place before the polls opened, then it would provide correct results, and only start producing altered ballots once the polls opened. This vector of attack is relatively easy to address, by having auditors only vote during open polling hours. However, there are more sophisticated ways of trying to determine when an auditor might be voting. For example, it is likely that if an auditor is using a script they will become exceptionally proficient at voting with that script, and are likely to move through the ballot very quickly. Malicious software could look for voting behavior patterns that matched this expected auditor voting pattern and produce "correct" ballots in these cases.

To defeat these kinds of attacks, audit team behavior would have to mimic actual voting patterns of real voters so the machines could not know that they were being audited. In unpublished work proposed by Wallach and Kortum, models of voting behavior could be developed based on actual user data at the individual and precinct level. These models would include likely candidate selections, predicted rolloff rates, time spent selecting the candidate for each race, time spent on the instructions, time spent on the review screen and other pertinent variables. Using these models, the voting behavior of a stereotypical voter could be randomly generated in real time in support of an audit operation. As an example, an auditor would step up to a voting machine and draw up a random stereotypical voter for that precinct that has been statistically determined. The auditor would then proceed to vote according to that profile. Since no two voting audit profiles would be identical, an attacker would have no way of knowing whether the voting machine was being audited or was being used by an actual voter.

While live audits may not be a panacea, they can certainly improve the security situation for BMDs, independent of whether or not voters successfully verify. Nonetheless, voters still provide the first line of defense and it is incumbent on the research community to continue to investigate the question of BMD verifiability and how to improve it.

## CONCLUSION

In conclusion, we believe this research has provided much greater clarity in our understanding of the problem at hand, suggesting a substantial change in the way we can attack the problem of potentially malicious BMDs. Our data strongly suggest that voters *can* detect changes, if they will only attempt to do so. This suggests that the first order of business in insuring BMD ballots are secure is more fully understanding how to motivate users to actually perform the auditing task of examining their own ballot for errors. This is a very different tack than has been taken in the past, when the assumption was that users simply are poor performers at this task. Instead, this appears to be true only if we aggregate the data by counting failure as unitary, when in fact there are two different ways voters can fail to verify: failing to check at all, and failing when they attempt to check. Performance on only one of those two is poor, which changes our understanding of the problem.

## ACKNOWLEDGMENTS

This work was supported in part by NSF grant numbers SMA-1853936 and SMA-1559393 and in part by a grant from the Williman and Flora Hewlett Foundation through the MIT Election Lab. We would also like to thank Ian Robertson for his assistance in running the experiments. The views and conclusions contained herein are those of the authors and should not be interpreted as representing the official policies or endorsements, either expressed or implied, of the NSF, the U.S. Government, the Williman and Flora Hewlett Foundation, MIT, or any other organization.